\documentclass[prd,nofootinbib,twocolumn,superscriptaddress,preprintnumbers,
               balancelastpage,floatfix]{revtex4-1}
\usepackage{graphicx}
\usepackage{slashed}
\usepackage{dcolumn}
\usepackage{bm}
\usepackage{natbib}
\usepackage{amsmath}
\usepackage{amssymb}
\usepackage{epstopdf}
\usepackage{hhline}
\usepackage{hyperref} 
\usepackage{color}
\usepackage{relsize}
 \usepackage[titletoc,toc,title]{appendix}
\usepackage{array}
\newcolumntype{P}[1]{>{\centering\arraybackslash}p{#1}}
\usepackage{multirow} 
\usepackage{chngcntr}
\usepackage[english]{babel}
\usepackage{romannum}

\newcommand{\aap}{{Astron.~Astrophys.}}

\newcommand{\mnras}{{Mon.~Not.~R.~Astron.~Soc.}}

\newcommand{\be}{\begin{equation}}
\newcommand{\ee}{\end{equation}}
\newcommand{\bea}{\begin{eqnarray}}
\newcommand{\eea}{\end{eqnarray}}

\definecolor{orange}{rgb}{1.0, 0.35, 0.21}

\newcommand{\antii}{Antilla~\Romannum{2} }
\newcommand{\craii}{Crater~\Romannum{2} }
\newcommand{\retii}{Reticulum~\Romannum{2} }

\begin{document}

\title{Diversity in density profiles of self-interacting dark matter satellite halos}
\preprint{TTK-19-17}

\author{Felix Kahlhoefer}
\email{kahlhoefer@physik.rwth-aachen.de }
\affiliation{Institute for Theoretical Particle Physics and Cosmology (TTK), RWTH Aachen University, D-52056 Aachen, Germany}

\author{Manoj Kaplinghat}
\email{mkapling@uci.edu}
\affiliation{Department of Physics and Astronomy, University of California, Irvine, California 92697, USA}

\author{Tracy R. Slatyer}
\email{tslatyer@mit.edu}
\affiliation{Center for Theoretical Physics, Massachusetts Institute of Technology, Cambridge, MA 02139, USA}
\affiliation{School of Natural Sciences
Institute for Advanced Study, Princeton, NJ 08540, USA}
\preprint{MIT-CTP/5117}

\author{Chih-Liang Wu}
\email{cliang@mit.edu}
\affiliation{Center for Theoretical Physics, Massachusetts Institute of Technology, Cambridge, MA 02139, USA}
\affiliation{School of Natural Sciences
Institute for Advanced Study, Princeton, NJ 08540, USA}

\begin{abstract} 
We present results from N-body simulations of self-interacting dark matter (SIDM) subhalos, which could host ultra-faint dwarf spheroidal galaxies, inside a Milky-Way-like main halo. We find that high-concentration subhalos are driven to gravothermal core collapse, while low-concentration subhalos develop large (kpc-sized) low-density cores, with both effects depending sensitively on the satellite's orbit and the self-interaction cross section over mass $\sigma/m$. The overall effect for $\sigma/m \gtrsim 3 \ \rm cm^2/g$ is to increase the range of inner densities, potentially explaining the observed diversity of Milky Way satellites, which include compact systems like Draco and Segue~1 that are dense in dark matter, and less dense, diffuse systems like Sextans and Crater~II. We discuss possible ways of distinguishing SIDM models from collisionless dark matter models using the inferred dark matter densities and stellar sizes of the dwarf spheroidal galaxies.
\end{abstract}

\maketitle

\noindent 
\section{Introduction}
Self-interacting dark matter (SIDM) was first proposed by ref.~\cite{Spergel:1999mh} as a way to explain the lower-than-expected dark matter (DM) densities in some galaxies. The idea was rediscovered about a decade ago when it was realized that DM self-interactions with observational consequences are generic in particle physics models with a hidden sector~\cite{ArkaniHamed:2008qn,Feng:2009mn,Feng:2009hw,Buckley:2009in}.  This led to a renewed interest in simulating SIDM halos~\cite{Vogelsberger:2012ku,Rocha:2012jg}, and a critical reevaluation of the astrophysical constraints~\cite{Peter:2012jh}. These efforts were further motivated by the small-scale structure puzzles, namely the too-big-to-fail problem~\cite{BoylanKolchin:2011de,BoylanKolchin:2011dk} and the inferences of low DM density cores in {\em some} galaxies~\cite{Moore:1994yx,Flores:1994gz,Simon:2004sr,Salucci:2007tm,KuziodeNaray:2007qi,deBlok:2009sp,Oh:2008ww,Salucci:2011ee}. Furthermore, SIDM can potentially explain the diverse rotation curves seen in galaxies~\cite{Ren:2018jpt,Kamada:2016euw}, which are not predicted by CDM simulations~\cite{Oman:2015xda}.

There is no concrete upper bound on the cross section at velocities relevant for dwarf galaxies ($\lesssim 50 \rm km/s$). The self-interaction cross section over mass of the DM particle, $\sigma/m$, could be as large as $50 \ \rm cm^2/g$ in these systems~\cite{Elbert:2014bma}. Such large cross sections are however strongly incompatible with the inferred central densities of galaxy clusters~\cite{Kaplinghat:2015aga,Bondarenko:2017rfu,Robertson:2018anx} as well as with bounds from major mergers~\cite{Wittman:2017gxn} and bright central galaxy wobbles~\cite{Kim:2016ujt,Harvey:2018uwf}. On the other hand, values of $\sigma/m$ below about $0.5 \ \rm cm^2/g$ do not seem to have a significant impact on the internal structure of dwarf spheroidal galaxies (dSph) of the Milky Way (MW)~\cite{Zavala:2012us,Valli:2017ktb}. These considerations highlight the importance of considering the velocity dependence of the self-interaction cross section~\cite{Feng:2009hw,PhysRevLett.106.171302,Loeb:2010gj,Tulin:2012wi,Aarssen:2012fx}. Indeed, a large cross section $\sigma/m \gg 1 \ \rm cm^2/g$ for velocities below about $50 \ \rm km/s$ that falls to about $0.1 \ \rm cm^2/g$ at velocities larger than about $1000 \ \rm km/s$ is consistent with all data and shows great promise in solving the small-scale puzzles~\cite{Kaplinghat:2015aga}.

For models with large cross sections at velocities $\lesssim 50 \ \rm km/s$, DM halos may potentially undergo gravothermal core collapse~\cite{Balberg:2002ue}. It was recently pointed out~\cite{Nishikawa:2019lsc} that for $\sigma/m \gtrsim 5 \ \rm cm^2/g$ this effect can play an important role for the evolution of satellite galaxies of the MW and introduce a correlation of central density with pericenter distance. In this paper, we focus on the effect of self-interactions on the DM density profile of satellite galaxies. We perform N-body simulations for different orbits and initial conditions for satellite galaxies, up to $t_\text{age} = 10 \, \text{Gyr}$, corresponding to 5 -- 8 orbital periods. We expect baryonic feedback effects to be small in the systems we study due to their low stellar content~\cite{Bullock:2017xww}, with the evolution being dominated by DM self-interactions.
As we will see, large self-interaction rates can lead to a diverse range of central density profiles for satellite galaxies.

We outline our simulation setup in Sec.~\ref{sec:setup}, and describe and justify the simulations we perform in Sec.~\ref{sec:sims}. We discuss results in Sec.~\ref{sec:results} and compare to observed ultrafaint dwarf galaxies in Sec.~\ref{sec:obs}. We summarize our key results in Sec.~\ref{sec:summary} and conclude in Sec.~\ref{sec:conclusion}. Our appendices review the algorithm we use (App.~\ref{sec:algorithm}); validate the simulation approach (App.~\ref{sec:validation}); explore the effects of a disk, for both equatorial and inclined orbits (App.~\ref{sec:disk}); and study the effects of allowing subhalos to evolve outside the MW and develop a core prior to their infall (App.~\ref{sec:core}).

\section{Simulating self-interactions\label{sec:setup}}
We  implemented DM self-interactions in the publicly available code The Astrophysical Multipurpose Software Environment  ({\tt AMUSE}) \cite{Pelupessy:2013yqa} which interfaces with {\tt GADGET-2} \cite{doi:10.1111/j.1365-2966.2005.09655.x}. We followed the self-interaction prescription of ref.~\cite{Rocha:2012jg}. The basic features of the algorithm are reviewed in App.~\ref{sec:setup}. We have checked that the scattering rate in the simulation is consistent with the analytical estimate based on the local density and local velocity distribution (see App.~\ref{sec:validation}). 

We seek to investigate how the density profiles of satellite halos (subhalos) change as they orbit the main halo, and compare the results for the SIDM and CDM models. This necessitates the inclusion of a sufficiently large number of particles in the subhalo,  in order to resolve structure at small radii, and inadequate force-softening might lead to artificial tidal disruption \cite{vandenBosch:2017ynq}. Therefore, we simulate a satellite halo with $10^6$ particles~\cite{vandenBosch:2017ynq} and a gravitational softening length of 30 pc. We also test a smaller softening length (10 pc) for an isolated halo, and find the density evolution is similar.  
We use {\tt Rockstar} \cite{Behroozi:2011ju} and {\tt pynbody} \cite{2013ascl.soft05002P} to analyze the simulation data. 

The results shown in this paper include only self-interactions between subhalo particles, not host-host or host-subhalo scattering. Including self-interactions between host particles would not change the conclusions of this work because they do not affect the MW density profile beyond a few kpc~\cite{Kaplinghat:2013kqa,Robles:2019mfq}. Since the subhalos that we consider never come closer than about 25 kpc, there are essentially no differences between the SIDM MW halo and the CDM halo.  

Scattering between particles from the halo and particles from the subhalo, on the other hand, are not in general negligible, as they can lead to mass loss and heating of the subhalo. Both of these effects are approximately proportional to the total scattering probability $p = \Sigma \, \sigma / m$, where $\Sigma = \int \rho \, \mathrm{d}x \approx 5 \times 10^8 \, \mathrm{M_\odot / kpc^{2}}$ is the integrated density of the host halo along the path of the subhalo 
for the satellite galaxy orbits we consider later. 
Since the orbital velocity $v_\text{orb}$ of the subhalo is $\mathcal{O} (100) \, \text{km/s}$ and therefore much larger than the escape velocity $v_\text{esc}$ of particles bound to the subhalo ($\mathcal{O} (10) \, \text{km/s}$), most interactions will lead to particle expulsion. 
The constraints imposed on the scattering cross section at $\mathcal{O} (100) \, \text{km/s}$ by the interactions of the subhalo and parent halo DM particles, and the implications for the velocity dependence of the self-scattering cross section, will be investigated separately. Here we note that if $p \ll 1$, which translates to $\sigma / m \ll 10 \, \mathrm{cm^2 / g}$, then the subhalo will be safe from evaporation due to scattering with the parent halo DM particles.

Heating occurs whenever a collision does not lead to particle expulsion, such that the transferred momentum is stored in the subhalo. This can happen either for very low scattering angles, or if a scattered particle scatters again before leaving the subhalo. The probability of the former is suppressed relative to evaporation by a factor $v_\text{esc}^2 / v_\text{orb}^2$, the probability of the latter is approximately given by $\Sigma_\text{sub} \sigma / m$, where $\Sigma_\text{sub} \approx 10^7 \, \mathrm{M_\odot / kpc^{2}}$ is the surface density of the subhalo~\cite{Kahlhoefer:2013dca,Kummer:2017bhr}. Thus, the effect of heating can be neglected as long as $\sigma / m < 100 \, \mathrm{cm^2 / g}$.

As we will see below, the diversity of dwarf galaxies favors a self-interaction cross section that at first sight violates the evaporation bound. This tension can however be resolved if the self-interaction cross section depends on the relative velocity in such a way that scattering at low velocities is enhanced and subhalo-subhalo interactions occur with higher probability than subhalo-halo interactions~\cite{Loeb:2010gj,Tulin:2012wi}. In the following, when considering $\sigma / m > 3 \, \mathrm{cm^2 / g}$, we implicitly assume that the evaporation constraint is satisfied through a (mild) velocity dependence. This approach enables us to focus our attention on interactions between the DM particles inside subhalos and neglect interactions with the MW halo's DM particles.  

\section{Interplay of self-interaction and tidal effects}
\label{sec:sims}
Recent results~\cite{Kaplinghat:2019svz} show that the ultra faint galaxies with small pericenter show a large scatter in their central densities. With pericenter distances of 20 to 40 kpc~\cite{2018A&A...619A.103F,2018ApJ...863...89S}, the ultra-faint galaxies Segue~1 and \retii have high densities~\cite{2011ApJ...738...55M,2011ApJ...733...46S,2015ApJ...808..108W} while \antii and \craii are extremely under-dense~\cite{Caldwell:2016hrl,2018arXiv181104082T}. Our motivation in this work is to see if SIDM models with large cross section can reproduce this large scatter, and how those results compare with the CDM model predictions. If the high densities of Segue~1 and Reticulum~II arise from self-interactions, then it follows that galaxies like Draco and Ursa Minor (with pericenter distances around 30 kpc~\cite{2018A&A...619A.103F}) will also be in the core collapse phase and exhibit larger densities than a field halo of similar $V_{\rm max}$. 

Our main aim is to explore how the interplay of thermalization due to self-scattering, the orbit of the satellite, and its DM concentration at infall dictates its final DM density profile. 
To explore the non-linear coupling between thermalization of the subhalo DM particles and the tidal interactions it experiences in the MW, we have run the following simulations. \\
{\bf (1)} SIDM simulations of a field halo with mass $M_{200}=10^9 M_\odot$ for high and low concentrations ($c_{200}$) for $\sigma/m = 0, 3,10\ \rm cm^2/g$. Here $M_{200}$ is the mass enclosed within a radius $r_{200}$ such that the average density within $r_{200}$ is 200 times the critical density of the Universe at $z=0$. The initial density profile has the Navarro-Frenk-White (NFW) form $\rho(r) = \rho_s (r/r_s)^{-1}(1+r/r_s)^{-2}$, where $r_s$ and $\rho_s$ are the radial and density scales. The concentration of the halo is defined as $c_{200}=r_{200}/r_s$. These isolated halos were evolved for $10\, \rm Gyr$. 
{\bf (2)} SIDM simulations with initial density profiles set to the NFW form with $M_{200}=10^9 M_\odot$ for $\sigma/m = 0, 3,10\ \rm cm^2/g$, and then evolved in a MW potential with and without a disk for $10\, \rm Gyr$. The subhalos simulated here are appropriate hosts for ultra-faint dwarf spheroidals of the MW, while the classical dSphs should be hosted by larger mass subhalos~\cite{BoylanKolchin:2011de}. We have included $\sigma /m = 0$ (cold collisionless DM or CDM), which provides an interesting point of comparison, as we discuss later. The host halo has a mass of $M_{200} = 10^{12} \, \text{M}_\odot, r_s = 24.6 \, \text{kpc}$. We have $10^6$ host halo particles in the simulation, which is sufficient to resolve the structure of the halo at the pericenter distances for the orbits we use.
{\bf (3)} SIDM halos evolved outside the MW for $5\ \rm Gyr$ with cross sections $\sigma/m = 3,10\ \rm cm^2/g$ and then evolved in a MW potential with and without disk for $5\ \rm Gyr$, for a total evolution time of $10 \ \rm Gyr$.

Our work is complementary to that of ref.~\cite{Dooley:2016ajo}, which studied the average properties of SIDM subhalos for a variety of self-interaction models. That work characterized the mass loss rate of SIDM halos using N-body simulations and showed that the half-light radius of satellites grows and stellar stripping is enhanced for cored profiles. Our focus is on characterizing the extremes of satellite evolution in SIDM models. 

The key idea we focus on is that for large enough cross sections, subhalos can follow different evolutionary tracks -- some can core collapse, while others undergo core expansion. In fact, all SIDM halos will eventually go through both these phases -- a long phase of central density decreasing with time, reaching a minimum and then increasing leading to gravothermal collapse \cite{Balberg:2002ue,Koda:2011yb,Elbert:2014bma,Essig:2018pzq,Nishikawa:2019lsc}. The timescale for core collapse of field halos is ${\cal O}(100 t_0)$ with $t_0^{-1} = a (\sigma/m) v_0 \rho_s$ with $a = \sqrt{16/\pi}$, and $v_0 = \sqrt{4 \pi G  \rho_s r_s^2}$~\cite{Balberg:2002ue,Essig:2018pzq,Nishikawa:2019lsc}. The density scale is related to the concentration of the halo as $\rho_s \propto c_{200}^3/(\ln(1+c_{200})-c_{200}/(1+c_{200}))$. We see that $t_0$ is most sensitive to the concentration of the halos, which is higher on average for the subhalos that orbit closer to the center of the MW~\cite{Moline:2016pbm}. This opens up the possibility of core collapse in subhalos, while retaining the core expansion behavior in the field~\cite{Nishikawa:2019lsc}. 

Given the importance of the concentration of the halos (as discussed above in terms of the collapse time scale $t_0$), we considered two extremes. In both cases, we initialized halos with the NFW density profile; one halo had a high concentration $c_{200} = r_{200}/r_s \simeq 30 $ with virial mass and scale radius $M_{\text{s}} = 10^{9} \, \text{M}_\odot, r_s = 0.7 \, \text{kpc}$ respectively, and and the other had a concentration $c_{200} \simeq 15$ with $M_{s} = 10^{9} \, \text{M}_\odot, r_s = 1.4 \, \text{kpc}$. These concentrations are respectively about 0.25 dex high and 0.05 dex low compared to the median CDM halos at redshift $z = 0$ in the field~\cite{Dutton:2014xda}. Tidal stripping will reduce their mass and increase their concentration; the deeper a satellite is embedded in the parent halo, the higher its concentration is on average~\cite{Moline:2016pbm}. Note that this implies (given the $t_0$ dependence) that at fixed mass (or $V_{\rm max}$), subhalos with their larger concentration will core collapse faster than field halos.   

To explore the tidal evolution of satellite galaxies we consider two different eccentric orbits for satellites, which we label as ``orbit A'' (apocenter distance $ = 60$ kpc) and ``orbit B'' (apocenter distance $ = 100$ kpc), both with pericenter distances of close to $25$ kpc. The goal of this setup is to see the interplay between the effect of tides and the orbital timescale. We expect the differences between CDM and SIDM models to be amplified for orbit A given the larger number of pericenter passages. 

In addition to the eccentricity of the orbit, another important consideration is the infall time, which determines the time that a satellite has been evolving within the MW potential. 
This is relevant for all MW satellites, not just the ultra-faint dSphs. For example, satellites like Draco, which had their star formation turned off 10-12 Gyr ago must have fallen into the MW early, while Fornax with its later star formation turnoff must have fallen in a few Gyr ago~\cite{Weisz:2014qra}.
To test for the impact of the infall time on the evolution of the central density, we evolve an isolated halo for 5 Gyr before putting them on the orbits A and B. In these cases, the subhalos start with a significantly cored density profile. 

Finally, we consider the effects of including a disk in the main halo. The presence of such a disk can significantly affect the density profile of the subhalos ~\cite{Penarrubia:2010jk,2012ApJ...761...71Z,2014ApJ...786...87B,Robles:2019mfq} and their survival~\cite{2017MNRAS.471.1709G,Kelley:2018pdy,Robles:2019mfq}, especially those whose orbits bring them close to the halo center.
To make the main halo + disk configuration stable, in this case we model the particles in the main halo and disk as non-dynamical, so they simply source the potential. This is valid because we are already neglecting the evaporation effect from subhalo-halo interactions. We provide further details on the disk modeling in App.~\ref{sec:disk}. The orbits of subhalos we show here are in the plane of the disk; subhalos on orbits perpendicular to the disk are affected to a lesser degree (see Fig.~\ref{fig:diskinc}).
Thus the impact of the disk is expected to vary with orbital inclination, but to be roughly bracketed by our results with no disk and with orbits parallel to the disk. Note that the impact of the disk is milder for the high concentration subhalos, and the inclination of the orbits will be correspondingly less important. It has also been shown that the overall difference between a disk and a spheroid of stars of the same mass is small compared to the effect of adding the stellar potential~\cite{2017MNRAS.471.1709G}

Since our main aim is to decipher the impact of thermalization on orbits with small pericenter distances, we have not varied the pericenter distance. Satellites on pericenter distances smaller than $20\ \rm kpc$ are likely to be destroyed by the presence of the disk~\cite{Garrison-Kimmel:2017zes}. Satellites with pericenter distances larger than this value should show milder core collapse, tending to the field halo behavior at larger pericenter distances like those for 
the classical dSphs Fornax and Sextans~\cite{2018A&A...619A.103F}.

\begin{figure}[t]
\includegraphics[scale=0.20]{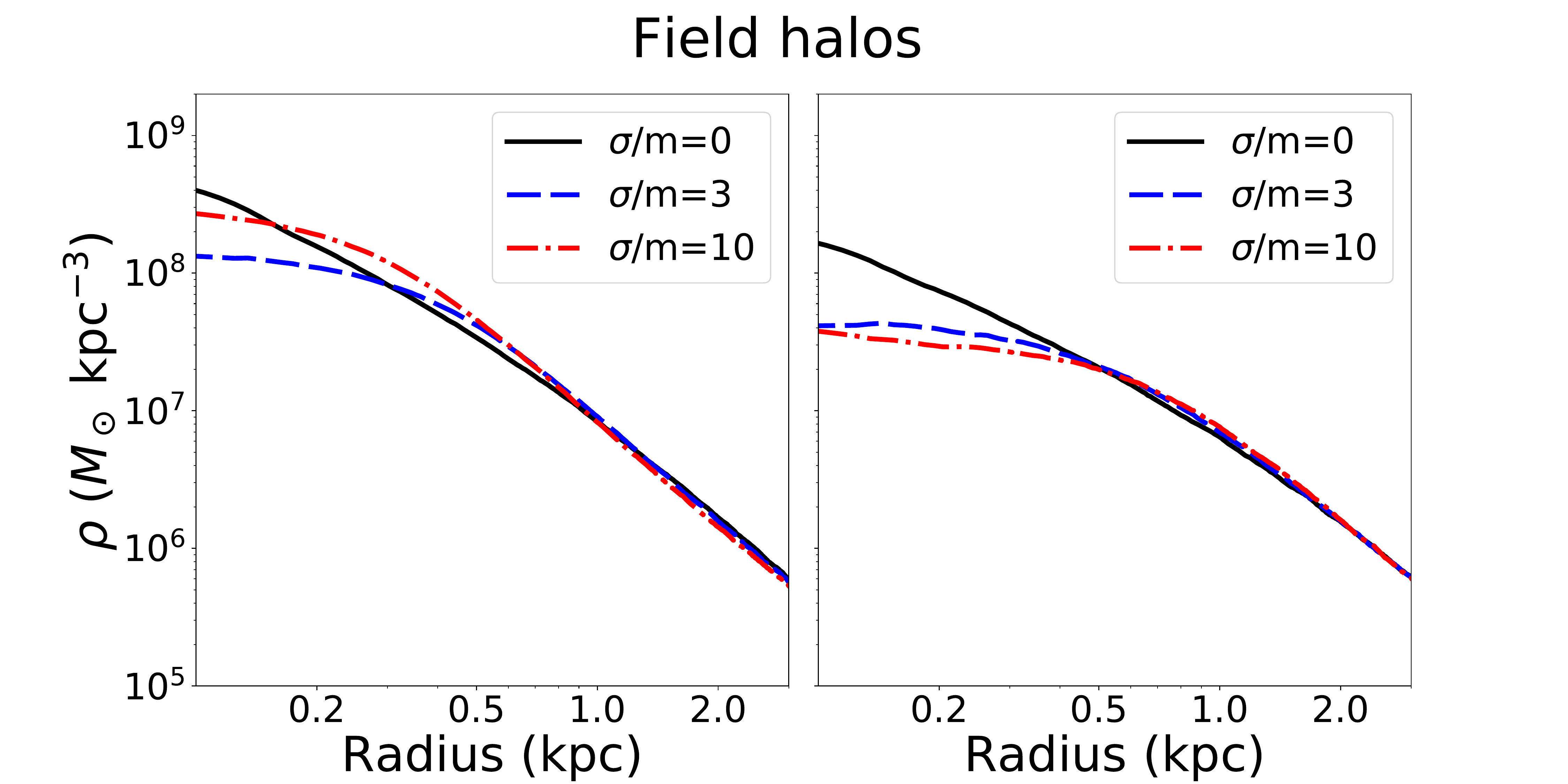}
\includegraphics[scale=0.20]{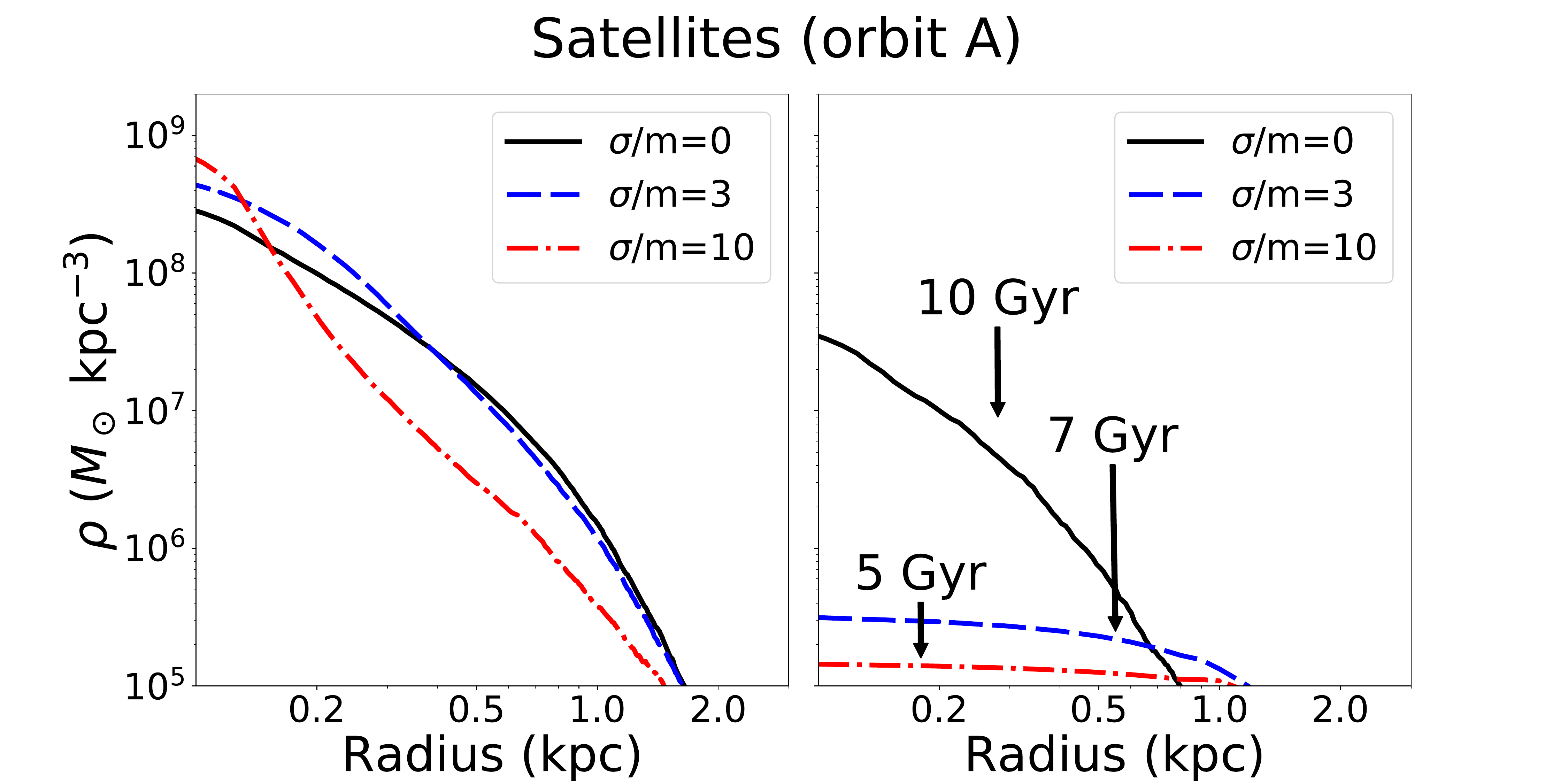}
\includegraphics[scale=0.20]{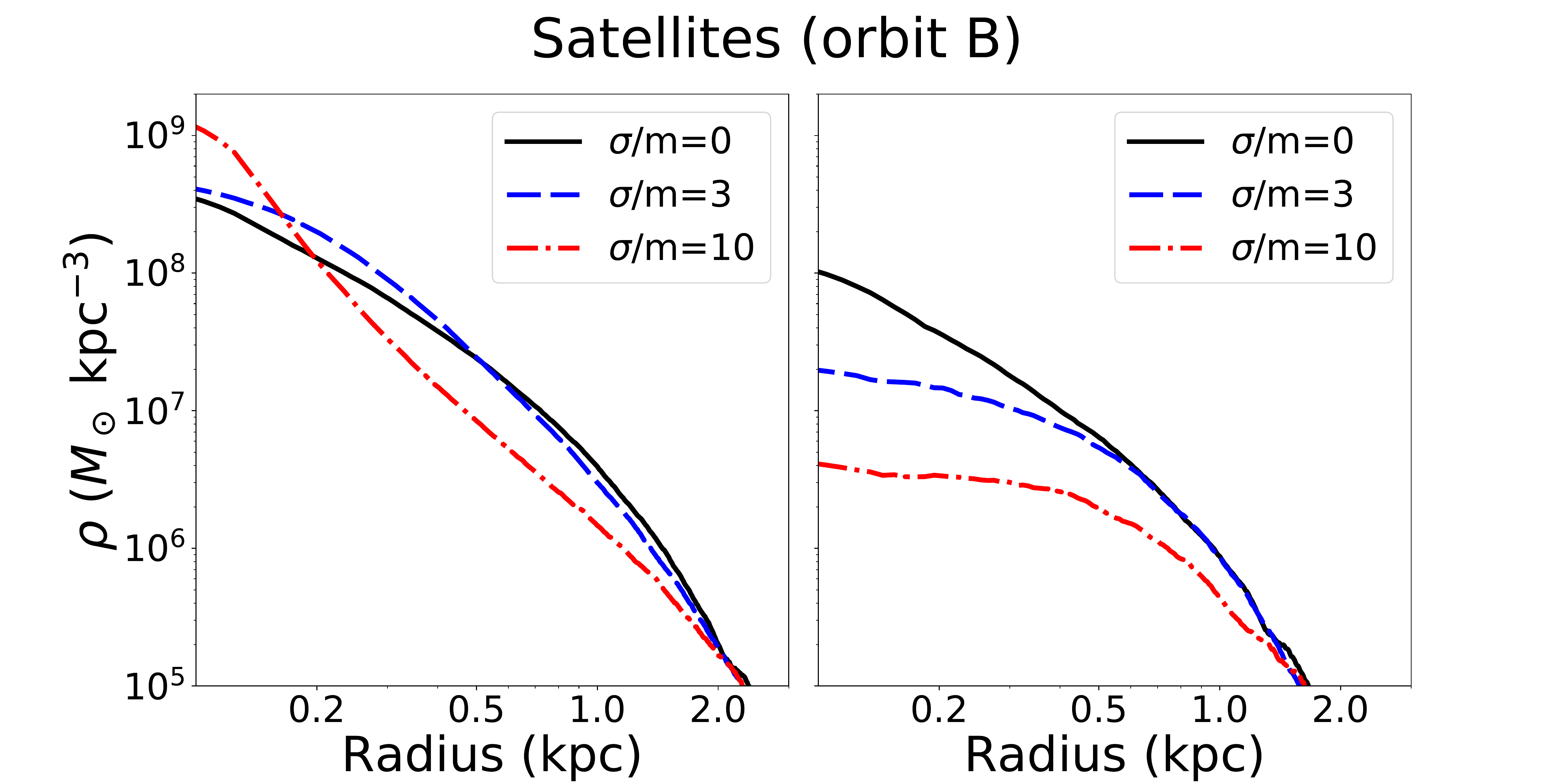}
\caption{The density profile of field halos (top) and satellites in shorter period orbit (middle) and longer period orbit (bottom) with high concentration (left) and low concentration (right), for self-interaction cross-sections of $\sigma/m=$0,3,10 cm$^2$/g. The evolution time is 10 Gyr except for the right middle panel, where we display the subhalo density profiles prior to their destruction.  }
\label{fig:isoall}
\end{figure}

\begin{figure}[t]
\includegraphics[scale=0.2]{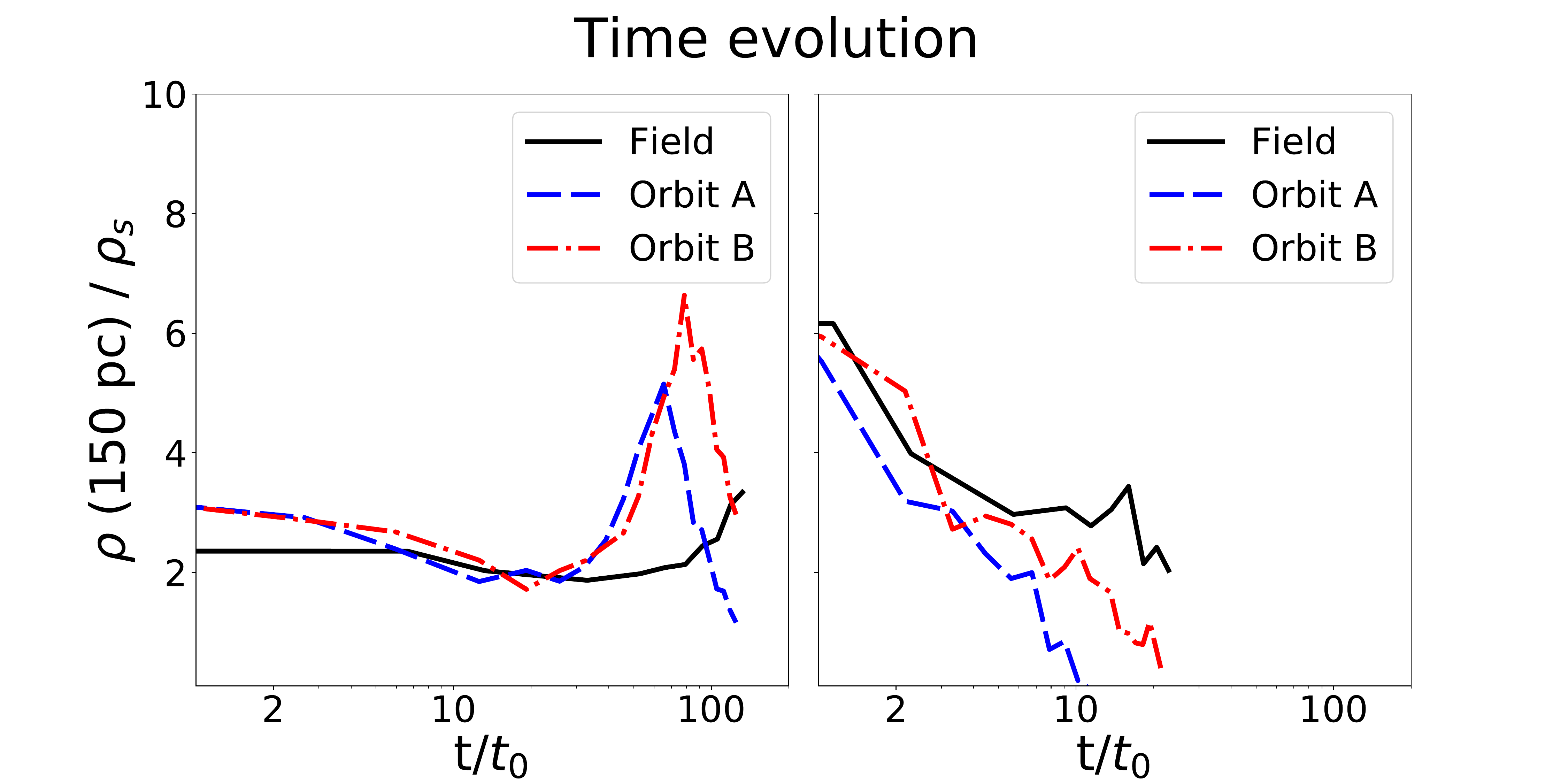}
\caption{The evolution of the dark matter density at 150 pc for the high concentration (left) and low concentration (right) subhalos on different orbits, assuming $\sigma/m = 10 \, \text{cm}^2/\text{g}$. The impact of the tidal interaction can be gauged by comparing these results with the evolution of an isolated halo (black line).}
\label{fig:test3denevo}
\end{figure}

\section{The large impact of small pericenter distances}
\label{sec:results}

Fig.~\ref{fig:isoall} shows the late-time density profile for the various simulations with and without self-interactions. It is instructive to first focus on the case including only gravitational effects, without self-interaction ($\sigma/m=0$). Fig.~\ref{fig:isoall} clearly demonstrates the low concentration subhalo (right panels) undergoes more tidal stripping than the high concentration one (left panels). It is also evident that subhalos moving on the shorter period orbit A (second row) feel stronger tides than those moving on orbit B (third row). 
The disk component plays an important role in enhancing the tidal effect, and the size of the effect has a strong dependence on orbital pericenter and the subhalo profile, as previously found in \cite{Penarrubia:2010jk}.  

When we add self-interaction between subhalo particles, Fig.~\ref{fig:isoall} reveals a richer range of behaviors. On the one hand, the high concentration subhalo grows more compact and undergoes gravothermal core-collapse so the core density profile is increased. Compared to the isolated case, the core density is significantly enhanced as predicted~\cite{Nishikawa:2019lsc}. This can be seen more clearly in Fig.~\ref{fig:test3denevo} where we show the evolution of the central density with time for $\sigma/m=10 \ \rm cm^2/g$.
In Fig.~\ref{fig:test3denevo} we have plotted the density at 150 pc in units of $\rho_s$ and time in units of $t_0$
The results of the left panel of Fig.~\ref{fig:test3denevo} agree qualitatively with studies of SIDM using a gravothermal fluid model \cite{Pollack:2014rja,Koda:2011yb,Nishikawa:2019lsc,Kummer:2019yrb}. Note that the density at 150 pc turns around for the high concentration subhalos and eventually starts to decrease. This is because the core is now at smaller radii and core collapse is dragging in material from larger radii (see Fig.~\ref{fig:isoall}), as predicted~\cite{Balberg:2002ue,Nishikawa:2019lsc}.

\begin{figure*}[t]
\includegraphics[scale=0.23]{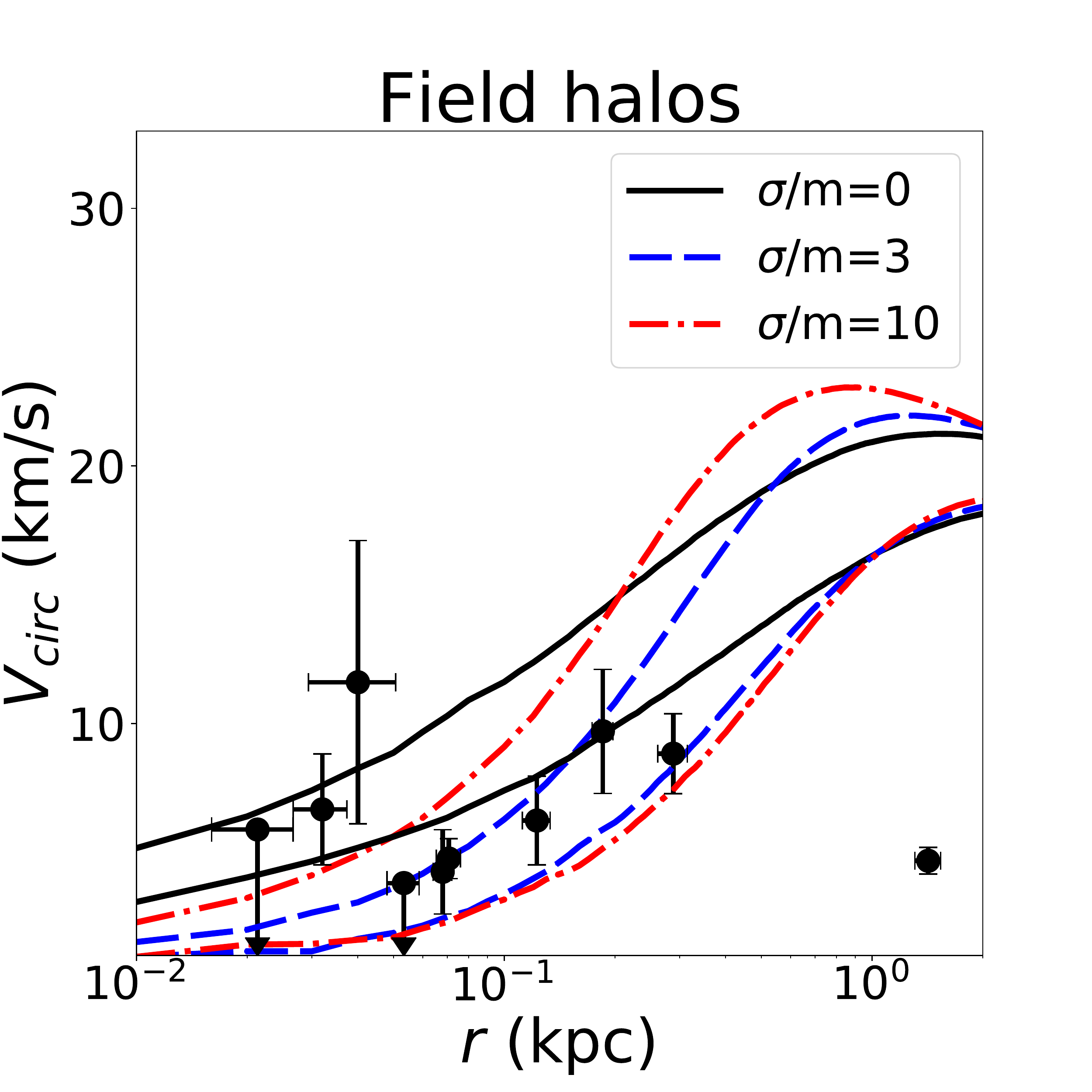} 
\includegraphics[scale=0.23]{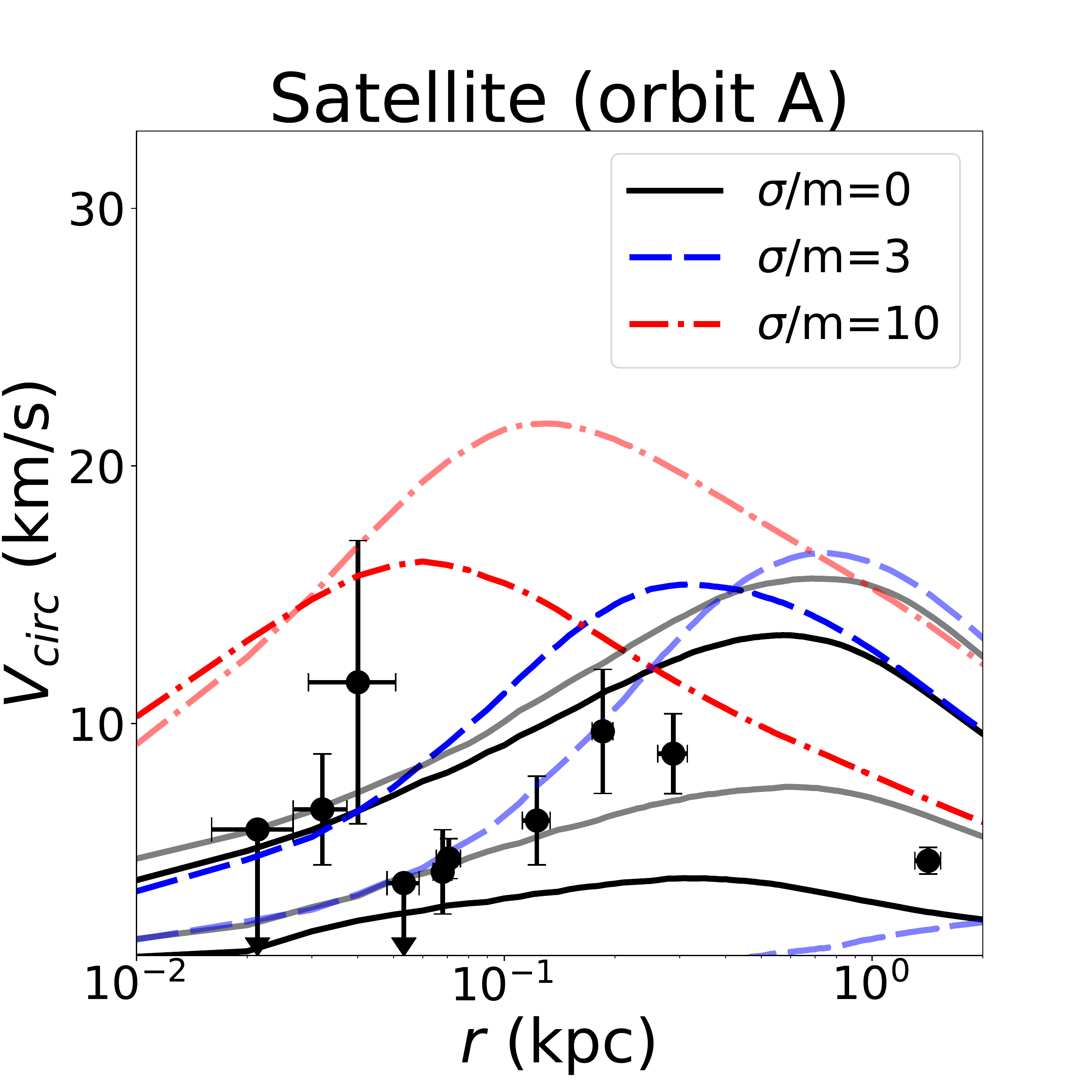}
\includegraphics[scale=0.23]{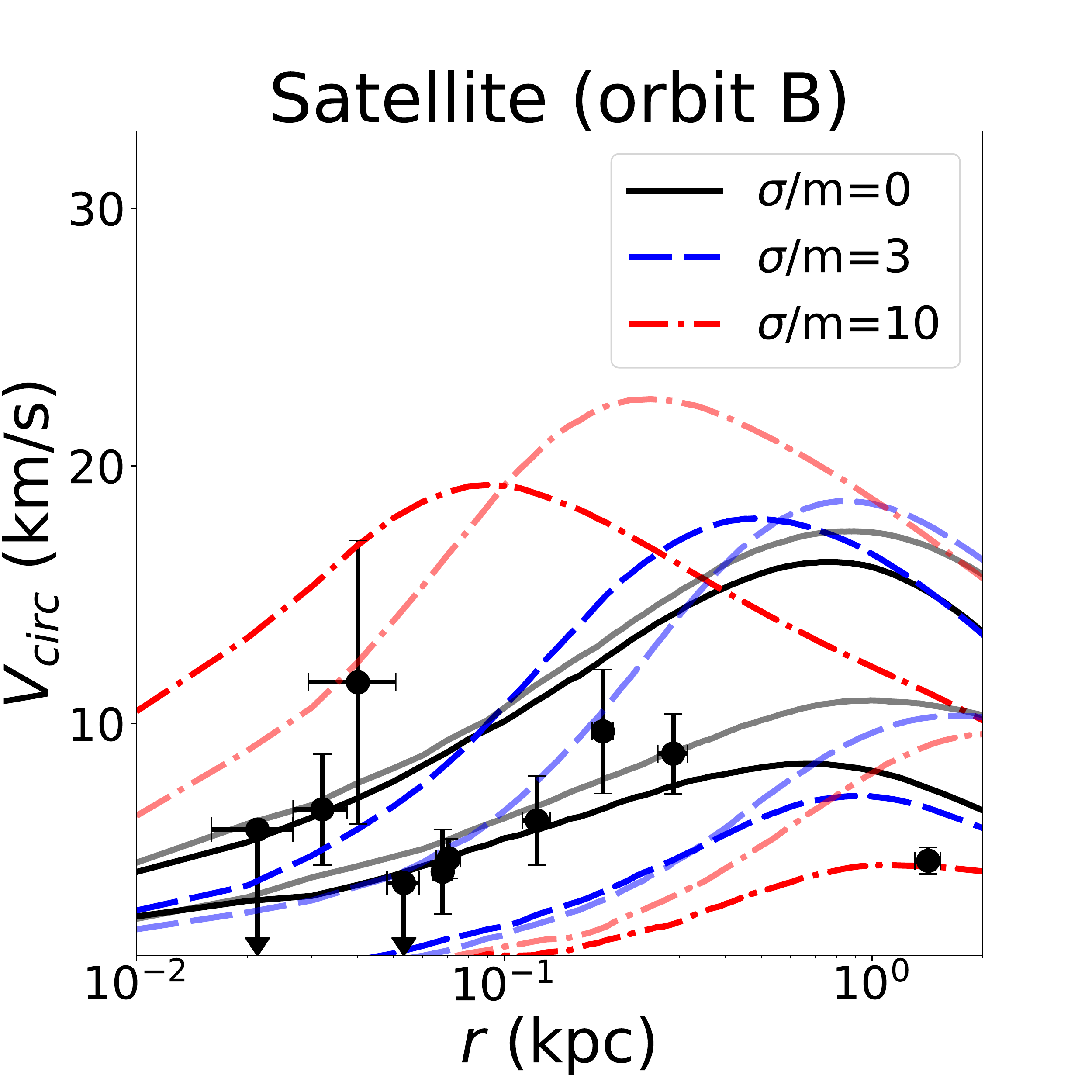} 
\caption{Characteristic radius and velocity of ultra-faint dwarf galaxies from our simulations and observational data. The black points indicate the half-light radius $r_{1/2}$ and circular velocity $V_{1/2}$ at that radius (or upper limits where applicable) obtained from observations. From left to right panels are isolated halos (left), satellite halos in orbit A (middle) and orbit B (right) in a Milky-Way like host halo for different self-interaction cross-sections. For the satellite halos, the darker color lines represent cases  with a NFW initial profile (early-infall) for low and high concentration, and the shallower color lines represent cored initial profiles (late-infall). The self-interaction cross-sections studied are $\sigma/m=0,3,10$ cm$^2$/g.}
\label{fig:obs}
\end{figure*}

The right panels of Fig.~\ref{fig:isoall} and Fig.~\ref{fig:test3denevo} shows that  self-interaction leads to a large cored profile in the low concentration subhalo for which the evolution time (set by $t_0$) is too long for core collapse to happen. The cored profile then leads to greater tidal stripping and a concomitant decrease in the central density, consistent with previous results~\cite{Dooley:2016ajo}. 
Eventually, the cored profile is mostly destroyed when the disk is present for $\sigma/m=10\ \rm cm^2/g$; the profile we show in Fig.~\ref{fig:isoall} is at 5 Gyr, when the subhalo is not yet destroyed. 

We observe a qualitatively similar range of late-time density profiles when the halo is allowed to evolve in isolation for 5 Gyr before beginning its orbit, mimicking a satellite that falls into the MW later. The profiles at infall are similar to those shown in the top panels of Fig.~\ref{fig:isoall}, since the differences in central density between 5 and 10 Gyr are small. The subhalo is allowed to orbit for another 5 Gyr, for a total evolution time of 10 Gyr. The final density profiles look similar despite the stark differences in the initial profile (see Fig.~\ref{fig:test3coreprofile} in App.~\ref{sec:core}). The low concentration subhalo on orbit A for $\sigma/m=10\ \rm cm^2/g$ survives, likely because it only has 5 Gyr evolution within the MW for the cored initial profile.

We have seen that the survival of the subhalo in the SIDM model depends crucially on the initial concentration, the cross section and the orbit. The range of subhalo densities recovered in our simulations argue that there must be some subhalos with a very low density like Crater~II~\cite{2016MNRAS.459.2370T,Caldwell:2016hrl} and Antilla~II~\cite{2018arXiv181104082T}. 
Both in CDM and SIDM with $3 \ \rm cm^2/g \lesssim \sigma/m \lesssim 10 \ \rm cm^2/g$, we see that the density within a kpc for the low concentration subhalo on orbit A (in the presence of the disk) has been significantly reduced; the same physics, including the impact of the inclination of the orbits, could explain the low DM density inferred for some dSphs. 

The key difference between CDM and SIDM is that the inner density profile for SIDM has a constant density core of almost 1 kpc. It has been shown that a dynamically sub-dominant stellar distribution will  expand in response to the expanding core size (decreasing core density)~\cite{Vogelsberger:2014pda}. This correlation between the stellar extent and the DM core provides a natural explanation for diffuse stellar components of galaxies like Crater~II and Antilla~II~\cite{2016MNRAS.459.2370T,2018arXiv181104082T}. Conversely, halos that have undergone core collapse, after significant tidal stripping, have small DM core sizes and high DM core densities. Interestingly, all of the ultra-faint dSphs that need high central densities also seem to have small half-light radii~\cite{Kaplinghat:2019svz}.

\section{Observations of ultra-faint dwarfs}
\label{sec:obs}
We can compare the range of density profiles obtained from our simulations to dwarf spheroidal galaxies with small pericenter distances. Given the mass of our simulated subhalos, we restrict our attention largely to the ultra-faint galaxies. These systems are DM-dominated and baryonic feedback is expected to be subdominant, so they are ideally suited for testing our scenario. 

The median values (or upper bounds) of the observational data are shown in Fig.~\ref{fig:obs}. To plot these points, we used the mass estimator at the half-light radius~\cite{Wolf:2009tu}. We include only dwarfs with pericenters smaller than 40 kpc inferred from stellar kinematics \cite{Kaplinghat:2019svz} or current distances smaller than 40 kpc. We exclude  Bo$\ddot{\text{o}}$tes \Romannum{2} because of large errors on velocity dispersion.  
The different curves in the plot are the circular velocity profiles of the simulated subhalos with $M_{200} = 10^9 M_\odot$ after 10 Gyr of evolution for different concentrations and infall times.
Surprisingly, all the models discussed, $\sigma/m=0,3,10 \ \rm cm^2/g$, seem to roughly get the right range of densities. For a more detailed comparison with observational data, one should perform cosmological simulations with a self-consistent model for populating satellites in subhalos~\cite{Kim:2017iwr}.

We urge caution in interpreting Fig.~\ref{fig:obs} given that we have not varied the mass of the subhalos or run a cosmological simulation to capture the distribution of subhalo properties or shown the errors in the mass estimates. What our work shows clearly is that it is not possible to explain the high densities of some of the ultra-faint dSphs in SIDM models without core collapse being important. Thus, we need $\sigma/m \gtrsim 3 \ \rm cm^2/g$ at relative velocities below about $30 \ \rm km/s$ for SIDM to be a viable explanation of galactic data. These larger cross sections also lead to significantly low central densities if the subhalos fell into the MW with low concentrations.

Note that for CDM, evolution within the MW halo serves to reduce the central DM density, while for SIDM the central density can increase or decrease depending on the initial concentration. For SIDM models with $\sigma/m \gtrsim 3 \ \rm cm^2/g$ we are able to identify distinct evolutionary paths for SIDM subhalos that lead to either core collapse or core expansion (to varying degrees) due to the combined effect of self-interactions and tides. 
The predicted scatter is larger for SIDM models with $\sigma/m \gtrsim 3 \ \rm cm^2/g$ than for CDM, and consistent with the data. Given the set of simulations and based solely on the measured densities at the half-light radii it is not possible to make an assessment whether $\sigma/m\simeq 0$ or $\sigma/m \gtrsim 3 \ \rm cm^2/g$ would be a better fit to the data. 

\begin{figure}[t]
\includegraphics[scale=0.27]{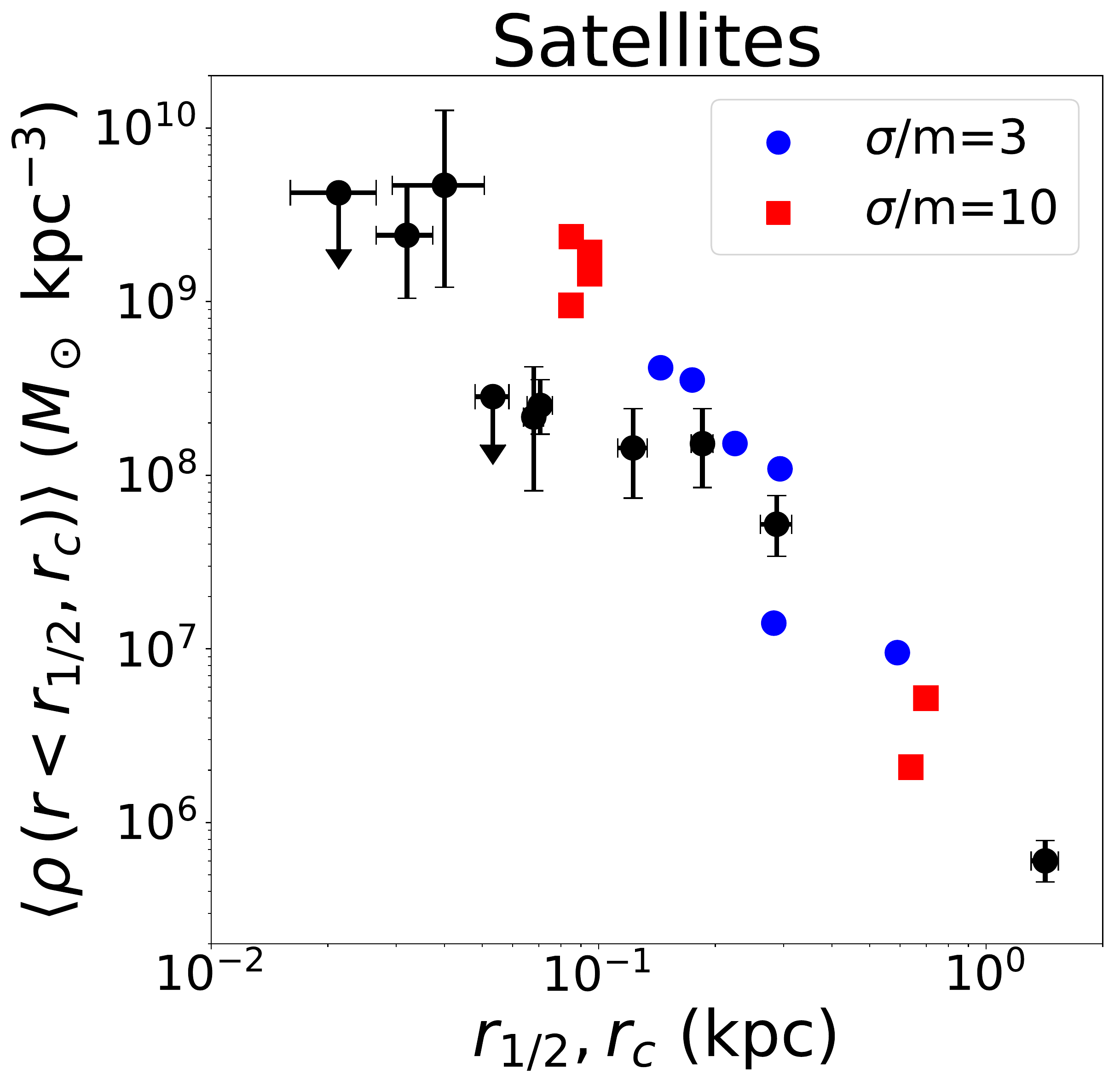} 
\caption{The mean density of the inner region for ultra-faint dwarfs. For observational data points (indicated by black points with error bars), the average density is defined from the mass enclosed within the half-light radius ($r_{1/2}$); for simulated data the mean density is evaluated within the core radius ($r_c$). The self-interaction cross sections are $\sigma/m=$ 3 (blue circle) and 10 (red square) cm$^2$/g.}
\label{fig:obs2}
\end{figure}

In identifying the subhalos that would host the observed ultra-faint dSphs, we notice that the stellar half-light radii should be correlated with the core sizes. This is because the least dense dSphs tend to be the ones with the large half-light radii and the most dense in DM are also compact in their stellar distribution.  We test this further in Fig.~\ref{fig:obs2} where we plot the mean density of our sample of ultra-faint dSphs with the half-light radius and compare that to the mean density of the subhalos within the DM core radius ($r_c$). The core radius here is defined as the radius where the density of the subhalo falls to half its maximum measured value in the simulation. For the high concentration subhalos, this could be an underestimate of the true core radius because of resolution issues. 

The comparison in Fig.~\ref{fig:obs2} is meant to illustrate the correlation between the core sizes in SIDM and stellar half-light radii and should not be interpreted strictly as a one-to-one mapping between these quantities.  We expect this correlation to form in SIDM as the stars respond adiabatically to the changing gravitational potential well. The data plotted in  Fig.~\ref{fig:obs2} show the trend of decreasing density with increasing half-light radii~\cite{Kaplinghat:2019svz} and the core densities from the simulated subhalos follow this trend. This comparison suggests that in order for SIDM to successfully explain the inferred densities of the ultra-faint dSphs, there must be a correlation between the stellar extent and the core size, similar to what is predicted for field halos~\cite{Vogelsberger:2014pda}. This could be a way to distinguish SIDM and CDM models through observations of ultra-faint dSphs. Within the SIDM model with $\sigma/m \gtrsim 3 \ \rm cm^2/g$, the large range of DM core sizes could also provide an explanation for the large range of observed stellar sizes, and this mechanism is distinct from processes discussed for collisionless DM~\cite{2018arXiv181202749W}.

It is plausible that further simulations could populate the region with even smaller central densities and larger core sizes. For example, subhalos which are disrupted due to the disk when on equatorial orbits (and hence do not appear on Fig.~\ref{fig:obs2}) may survive, albeit with large cores and low densities, on inclined orbits where the effect of the disk is reduced (see Fig.~\ref{fig:diskinc}). In this regard, we note that the inferred orbit of \craii is more eccentric~\cite{2018A&A...619A.103F} than the orbits we have simulated, and more work is required to validate our proposed explanation. 

In addition to the scatter in the predicted densities and the stellar sizes, SIDM and CDM models could be distinguished through measurements of the slope of the density profile of the ultra-faint dSphs. In order for SIDM to explain the low densities inferred for the diffuse dSphs, the core sizes have to be large. In CDM, the subhalos lose mass but the density profile stays cuspy. This is a clear prediction of SIDM that awaits future tests.

\section{Summary of results}
\label{sec:summary}
Our findings are the following. We have found that high concentration halos collapse faster than low concentration halos in the field, as expected from analytic arguments. A corollary of this result, which we do not explore here, is that we can use the distribution of density measurements in the field to place an upper limit on the cross section (i.e., leave enough galaxies with cores in the field). For subhalos on orbits with small pericenter distances (which suffer large mass losses), the differences between low and high concentration halos seen in the field are magnified. In CDM, both high and low-concentration subhalos suffer mass loss in the center. In SIDM with $\sigma/m >$ few $\rm cm^2/g$, high concentration subhalos core collapse and become denser than in CDM, while low-concentration suffer dramatic reductions in density. 

We demonstrated that these results are sensitive to the apocenter, with the effects being more pronounced for subhalos on smaller period orbits. Our MW model included a fixed disk and we showed that its presence amplifies the effects futher (because of the increase in tidal stripping). 
We found that our overall results are not very sensitive to the infall time for SIDM halos, except in the sense that there is less time for evolution in the field. The differences between starting out with a cusp (early infall) and large core (late infall) were muted compared to the other effects. 

The diversity in the simulated SIDM halos seems to be what is required to explain the full range of densities inferred in the MW satellites. Compare and contrast the cases of Crater II and Draco -- they both have similar pericenter distances~\cite{2018A&A...619A.103F}, similar luminosities (within a factor of 2) but they have very different half-light radii and very different DM densities~\cite{2019arXiv190105465S}. This has a natural explanation in SIDM models in terms of Draco forming in a high-concentration halo and Crater~II in a low concentration halo. Compact ultra-faint dSphs like Segue~1 would also form in high-concentration subhalos like Draco, but the subhalos would likely have lower masses similar to what we have simulated. The key difference between CDM and SIDM low-concentration halos that suffer significant tidal mass loss is that SIDM models unequivocally predict a large core for these subhalos, which likely host satellites like Crater~II and Antilla~II. 

We discussed the impact of the stars responding to the adiabatic changes in the DM gravitational potential of the subhalo. This physics correlates the sizes of the stellar systems and its DM density cores in SIDM models, which should be applicable to both the classical and ultra-faint dSphs. Those subhalos that undergo core collapse accelerated by the MW's tides will have high central DM densities and smaller core sizes: these subhalos seem to be the ones that are required to host compact satellites like Draco and Segue~1 in SIDM models. On the other hand, those that undergo core expansion due to self-interactions and tidal effects will have low central DM densities and large core sizes: diffuse satellites like Crater~II and Sextans would form this way in SIDM models. Thus, the large range of predicted core sizes in our work could provide a natural explanation for the large range of observed stellar sizes for the MW satellites. 

\section{Conclusions}
\label{sec:conclusion}

In this paper, we use N-body simulations to study the diverse evolutionary paths for satellite galaxies in models with self-interaction cross section per unit mass $\sigma/m \gtrsim 3 \ \rm cm^2/g$. Inclusion of self-interactions allows the density profile of subhalos to have richer diversity than in the case of collisionless cold DM, and the presence of tidal forces amplify the differences. The simulations indicate that various parameters, specifically the halo concentration, orbital eccentricity and the orbital period, strongly affect the DM density profile of the satellites. Our results, both in the context of SIDM and CDM models, provide an avenue to explain the diversity of densities seen in ultra-faint dwarf spheroidal galaxies of the MW. 

We conjectured that the stellar extent of the Milky Way satellites would be correlated with the DM core sizes of the subhalos in SIDM models, which could provide an explanation for the large range of observed stellar half-light radii. Comparing the range of simulated subhalos to the inferred densities for the ultra-faint dwarfs suggests that $\sigma/ m \gtrsim 3 \ \rm cm^2/g$ at collision velocities less than about $30 \ \rm km/s$ is required for SIDM models, which agrees well with the analysis of field galaxies~\cite{Ren:2018jpt} where  collision velocities of 30--300 km/s are accessible.

\emph{Note added:} During the completion of this work, a related study \cite{Sameie:2019zfo} appeared; our results are complementary, as we focus on ultra-faint dwarf galaxies, while that paper concentrates on the potential of SIDM models to explain brighter dwarfs such as Draco and Fornax.
\bigskip

\noindent 
{\bf Acknowledgments.}
We thank Kimberly K. Boddy, Kai Schmidt-Hoberg, Mauro Valli and Hai-Bo Yu for useful discussions.
FK acknowledges support from the Deutsche Forschungsgemeinschaft (DFG) through the Emmy Noether Grant No.\ KA 4662/1-1 and the Collaborative Research Center TRR 257 ``Particle Physics Phenomenology after the Higgs Discovery''. TRS and CLW are supported by the Office of High Energy Physics of the U.S. Department of Energy under grant Contract Numbers DE-SC00012567 and DE-SC0013999. CLW is partially supported by the Taiwan Top University Strategic Alliance (TUSA) Fellowship. TRS is partially supported by a John N. Bahcall Fellowship. MK is supported by the NSF Grant No.~PHY-1620638.

\bibliography{dwarf}

\clearpage
\newpage
\onecolumngrid

\begin{appendix}


\section{Review of the simulation algorithm}
\label{sec:algorithm}

Assuming a self-interaction cross-section per unit mass of $\sigma/m$, the probability for scattering off a particle $j$ with mass $M_j$, per particle $i$ with relative velocity $\mathbf{v}$ is given by:
\begin{equation}
P\left(j | i\right) = \dfrac{\sigma}{m} M_j |\mathbf{v}|   g_{ji}
\end{equation}
where the final factor $g_{ji}$ takes the smoothing kernel $W(r,h)$ into account, and is given by
\begin{align}
g_{ji} = \int_0^{max(h_{si},h_{sj})}  d^3 \mathbf{x'} W\left(|\mathbf{x'}|,h_{si}\right)W\left(|\delta \mathbf{x}_{ji}+\mathbf{x'}|,h_{sj}\right).
\end{align}
We adopt a spline kernel, the same as used in \texttt{GADGET-2}. The self-interaction softening length $h_{si}$ in principle can be different from the one used in gravitational interactions, and the size may depend on the typical density of the simulated objects, so one needs to check for convergence. The total probability of interaction between two equal-mass objects scattering is symmetrized to be
\begin{equation}
P_{ij} = \dfrac{P\left(i|j\right)+P\left(j|i\right) }{2}.
\end{equation}
Notice that this symmetrization assumes all simulation particles have the same mass; we will consider the scattering of unequal mass particles in a companion paper. The time-step is adaptive, so it is decreased by a factor of 2 when $P_{ij}$ is larger than $P_\text{max}$: we choose $P_\text{max} = 0.2$.  The velocity change when two DM particles interact is modeled by elastic scattering: 
\begin{align}
\mathbf{v}'_i &= \mathbf{V_c} + \dfrac{m_j \,  |\mathbf{v}|}{m_i+m_j} \mathbf{e} \, , &
\mathbf{v}'_j &= \mathbf{V_c} - \dfrac{m_i \, |\mathbf{v}|}{m_i+m_j} \mathbf{e} \, ,
\end{align}
where $\mathbf{V_c}$ is the center of mass velocity, and the scattering angle $\mathbf{e}$ is a unit vector in the scattering direction. 

\section{Validation of the simulation setup}
\label{sec:validation}

As a demonstration and test of our simulation setup, we track the self-interaction rate for an isolated halo. We use the low concentration subhalo in this paper as an example: the virial mass, scale radius $M_{200}=10^{9} \, \text{M}_\odot, r_s = 1.4 \, \text{kpc}$ respectively, and the halo contains $10^6$ particles. 

\begin{figure}[b]
\includegraphics[scale=0.27]{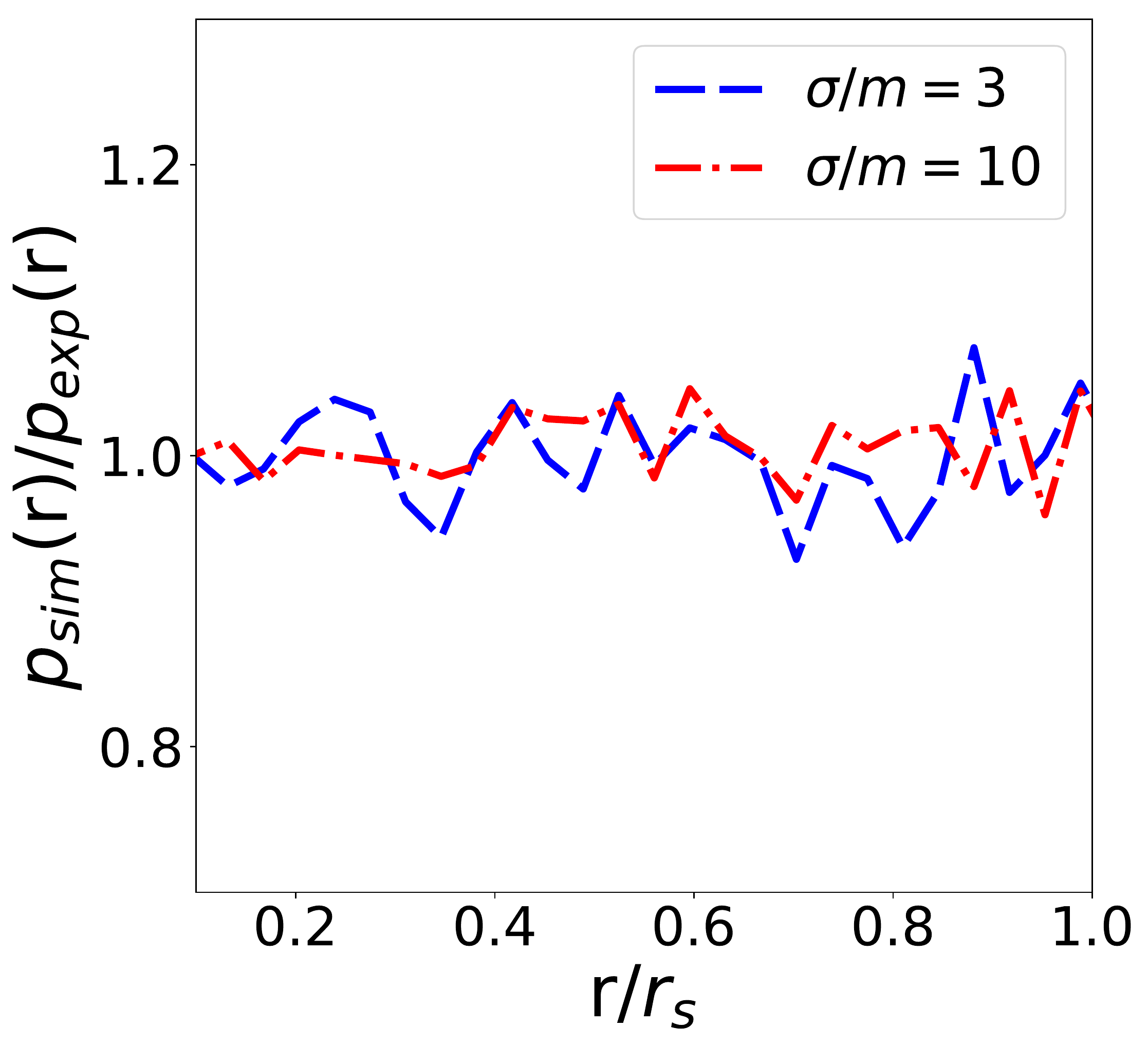}
\includegraphics[scale=0.27]{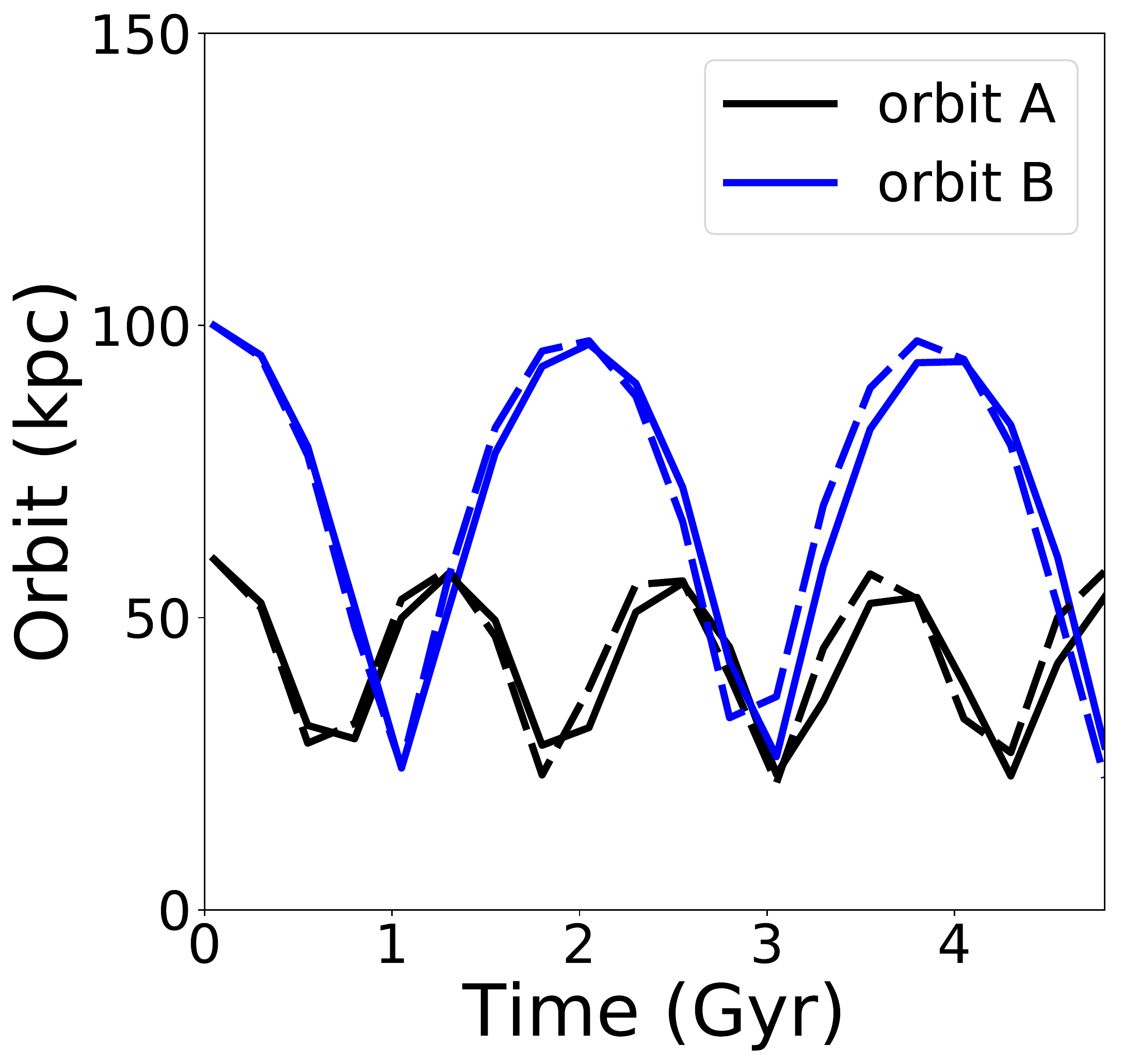}
\caption{Left panel: The ratio of probability of scattering in a simulation $p_{\text{sim}}$(r) compared to an analytic estimate $p_{\text{exp}}$(r) for a field halo, as a function of radius radius $r$, over 1 Gyr, assuming two different self-interaction cross-sections ($\sigma/m = 3$ cm$^2$/g, 10 cm$^2$/g).  Right panel: The simulated orbits of satellite galaxies relative to the center of the main halo, with apocenters at 100 kpc (black) and 60 kpc (blue). The solid (dotted) line indicates whether the main halo is with (without) a disk.}
\label{fig:test1_prob}
\end{figure}

The gravitational softening length is set to be $30\, \text{pc}$ and the ratio of the self-interaction softening length $h_{si}$ to the gravitational softening length is 0.25. This softening length has been confirmed to be sufficient in \cite{Rocha:2012jg}, when considering scatterings of target particles on background particles. We show it is also sufficient when considering isolated halos for a range of self-interaction cross-sections. 

We compare the probability of scattering for each radial bin from simulations, $p_{\text{sim}}$(r,t), with the expected number of scattering events from an analytic estimation, $p_{\text{exp}}(r,t)$, which during time interval $\delta t$ is:
\begin{equation}
p_{\text{exp}}(r,t) = \dfrac{\langle \sigma v_\text{rel} \rangle}{m}   \rho(r,t)  \delta t \sim  \dfrac{4}{\sqrt{\pi}} \dfrac{\sigma}{m} \rho(r,t) \sigma_v(r,t)   \delta t,
\end{equation}
where we assumed the velocity follows a Maxwell distribution with one-dimensional velocity dispersion $\sigma_v$. Summing over interactions during 1 Gyr, the ratio of the two probabilities is shown in the left panel of Fig.~\ref{fig:test1_prob}. The results are consistent with each other for the range of cross-sections we consider in this paper.


\section{Supplemental information for modeling of a Galactic disk and orbit}
\label{sec:disk}

The disk density in our simulations falls exponentially in radius $R$ and follows a $\text{sech}^2$ function in height $z$. We take a thin disk model where the mass, scale length, and scale height are MW-like with $M_{d} = 3.5 \times 10^{10} \, \text{M}_\odot, R_d = 2.6 \, \text{kpc}$, and $z_d = 300 \, \text{pc}$ \cite{doi:10.1146/annurev-astro-081915-023441}. The value of $z_d$ is extracted from a fit with an exponential model but the difference with sech$^2$ is small for $z> z_d$. The orbits we consider with and without disk are shown in the right panel of Fig.~\ref{fig:test1_prob}.
As expected, the disk effect on the orbit itself is small because the orbit pericenter is $\sim 25$ kpc, much larger than the disk radius.

\begin{figure}[b]
\includegraphics[scale=0.27]{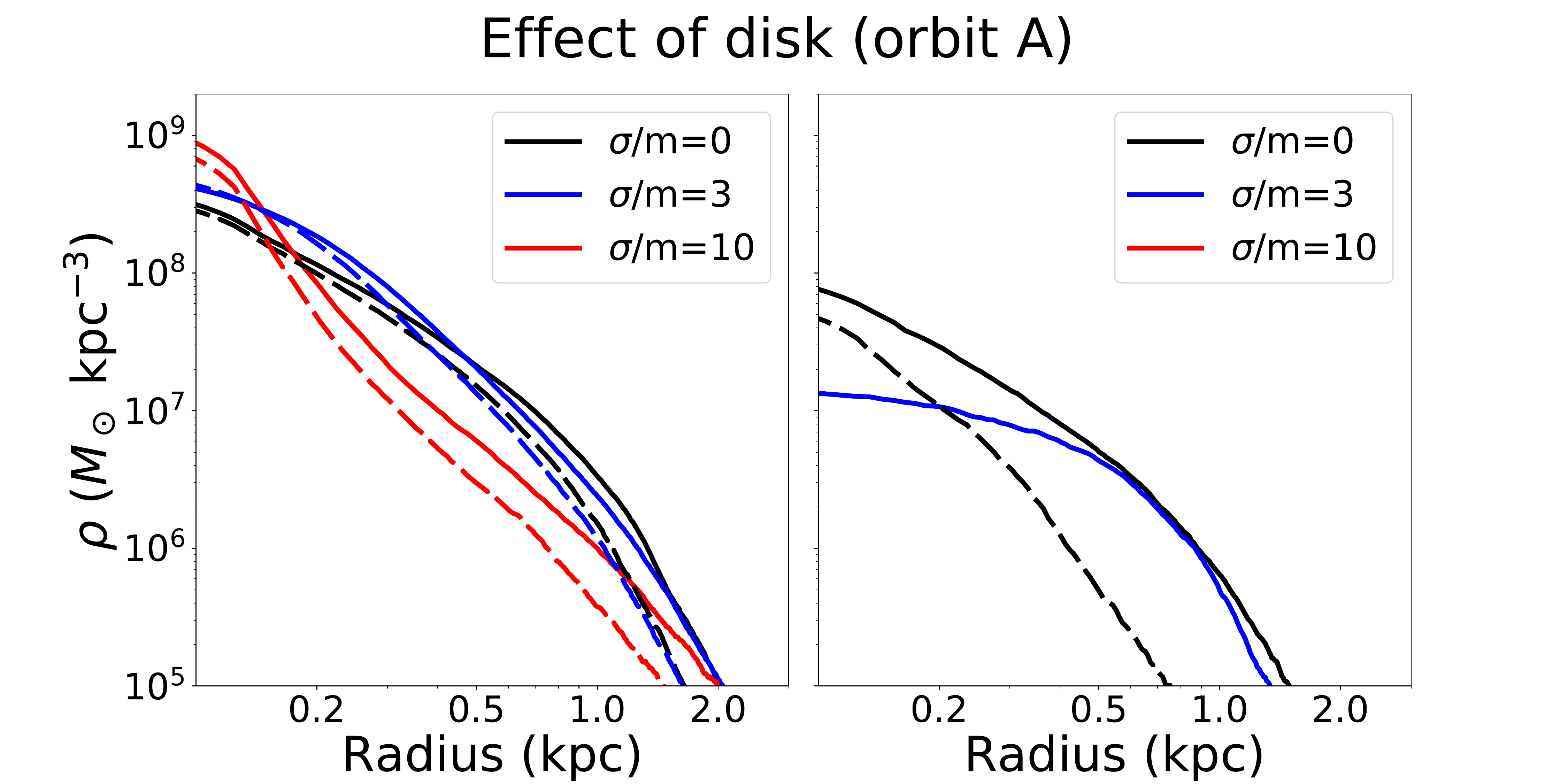}\\
\vspace{0.5cm}
\includegraphics[scale=0.27]{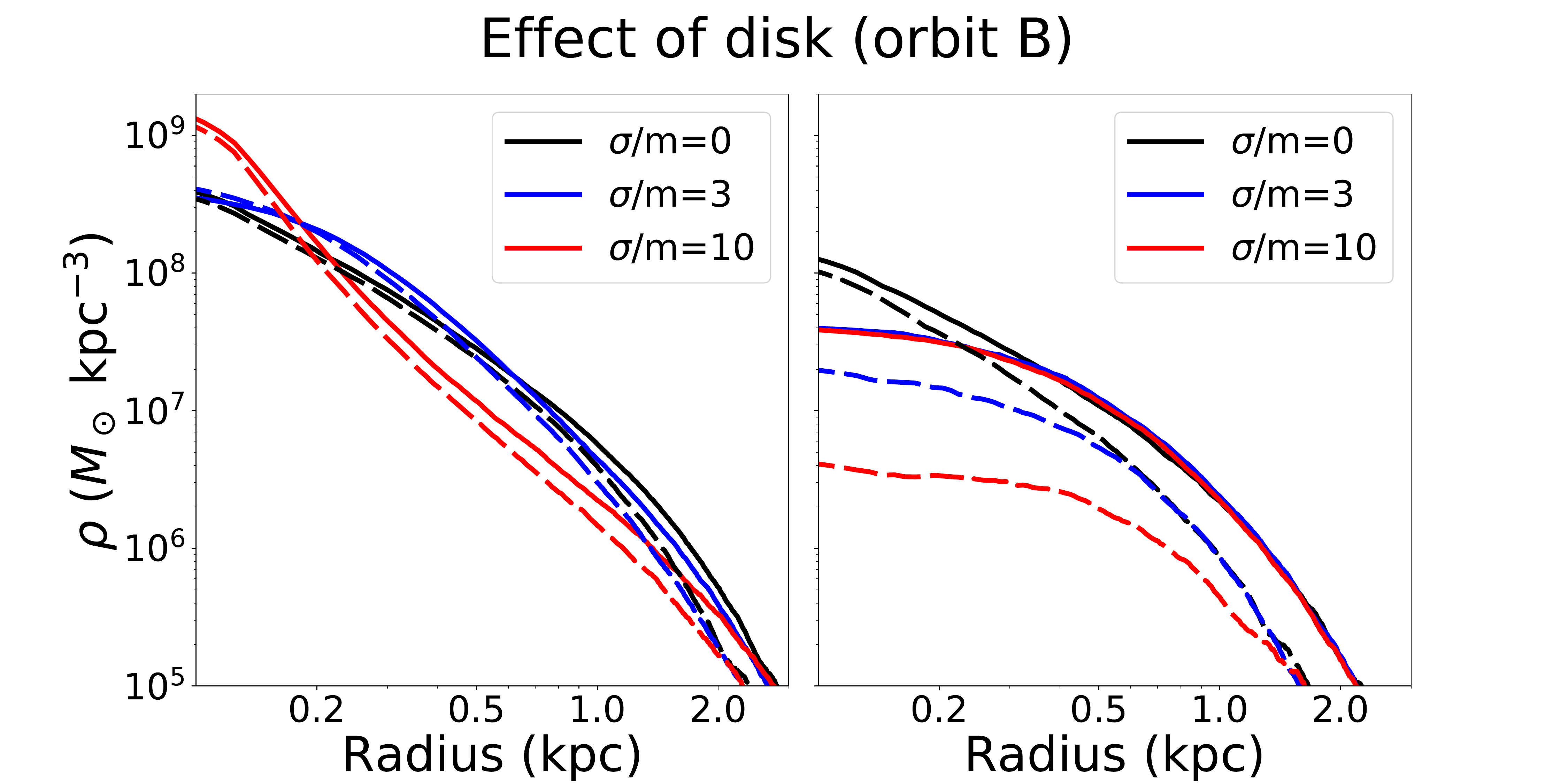}
\caption{The tidal effect from host halo with and without disk on high (left) and low (right) concentration halos for different orbits (solid: without disk, dashed: with disk). The curves not shown in the panel indicate the subhalo has been tidally destroyed at late time. The self-interaction cross sections tested are $\sigma/m=0,3,10$ cm$^2$/g.}
\label{fig:diskprofile}
\end{figure}

The effect of the disk on different concentrations, orbits and self-interaction cross section is shown in Fig.~\ref{fig:diskprofile}. For the high concentration halos, the central region is resilient to tides so the disk does not play a significant role. However, the low concentration subhalos are more vulnerable in the presence of the disk and the cored profile from self-interaction makes them even more sensitive to the disk.

\begin{figure}[h]
\includegraphics[scale=0.27]{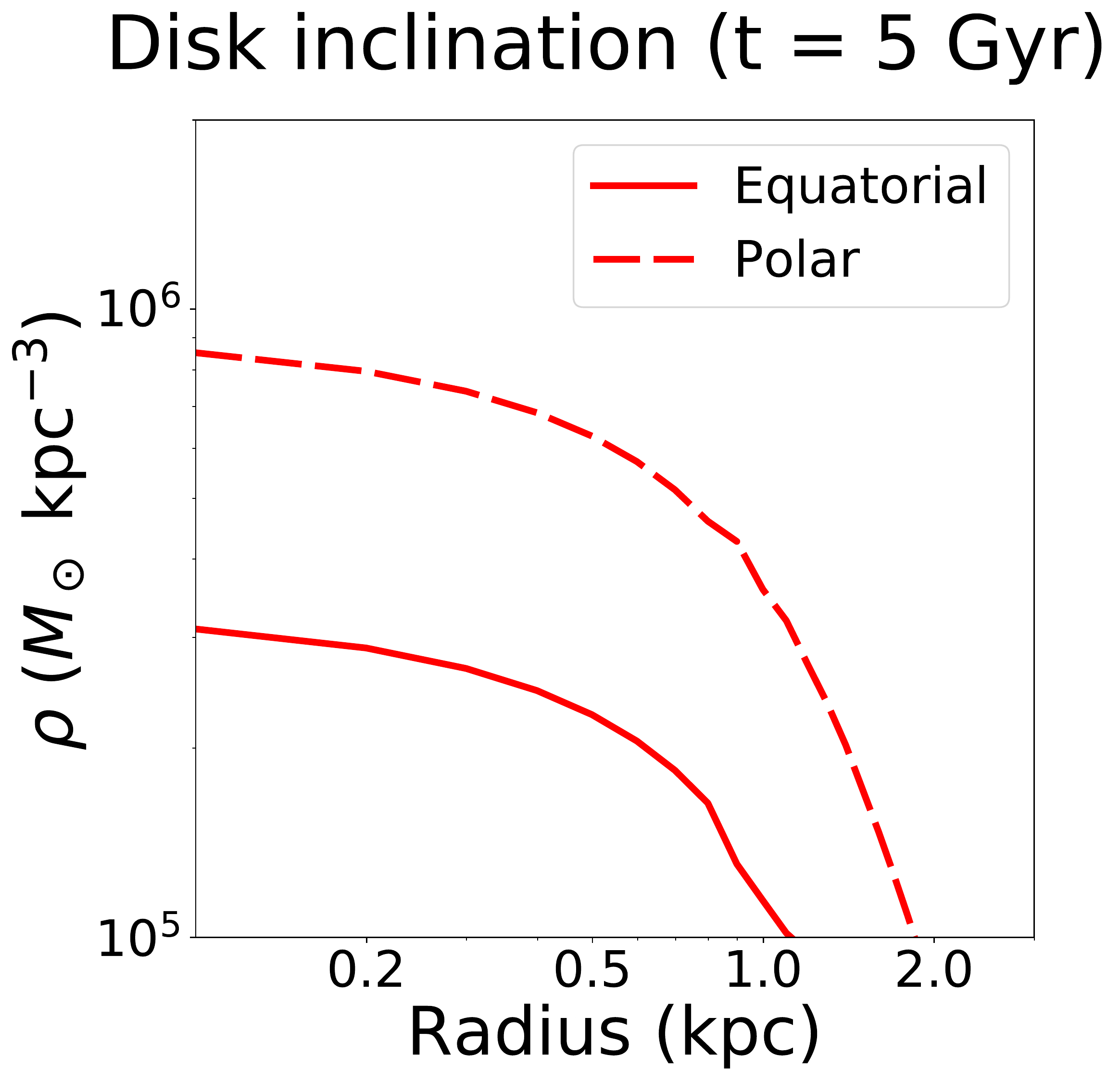}
\caption{The density profile of a low concentration subhalo at evolution time 5 Gyr (before the subhalo is largely destroyed), assuming $\sigma/m = 10 \, \text{cm}^2/\text{g}$, for the equatorial (parallel to disk) and polar (perpendicular to disk) inclinations of Orbit A.}
\label{fig:diskinc}
\end{figure}

Notice the results in Fig.~\ref{fig:diskprofile} are for equatorial orbits (in parallel to the disk). To test the effects of different orbital inclinations, we also run a simulation on the subhalo suffering the largest effect from the disk -- i.e. a low concentration subhalo in Orbit A with $\sigma/m = 10 \, \text{cm}^2/\text{g}$, with the orbit now chosen to be polar (perpendicular to the disk). Fig.~\ref{fig:diskinc} shows that in this example, with the largest effect from the disk, the central density profile of the subhalo in an equatorial orbit is smaller by a factor of three compared with the one in a polar orbit. 

\section{Effect on halo density evolution of varying infall times}
\label{sec:core}

\begin{figure}[h]
\includegraphics[scale=0.27]{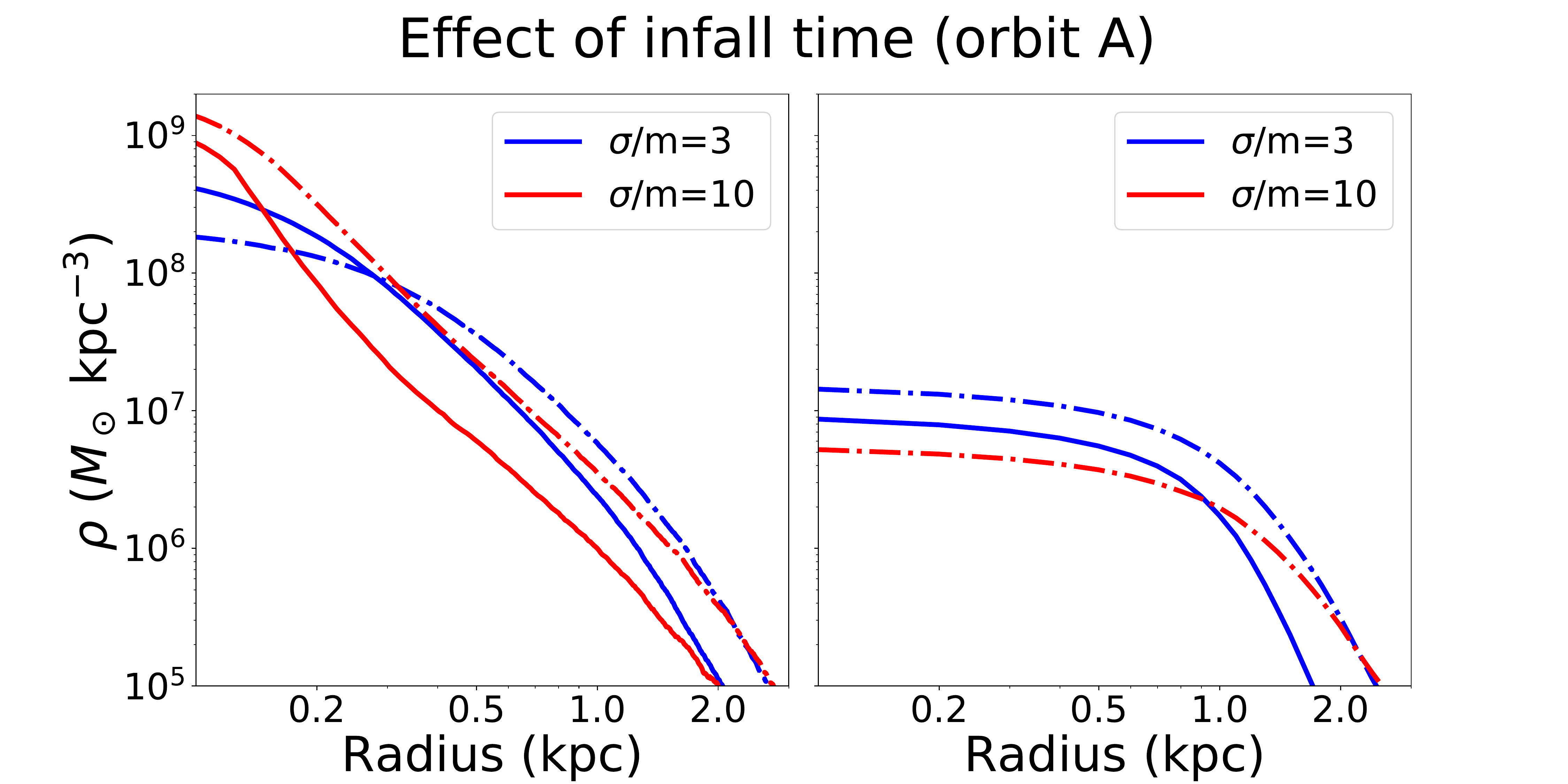}\\
\vspace{0.5cm}
\includegraphics[scale=0.27]{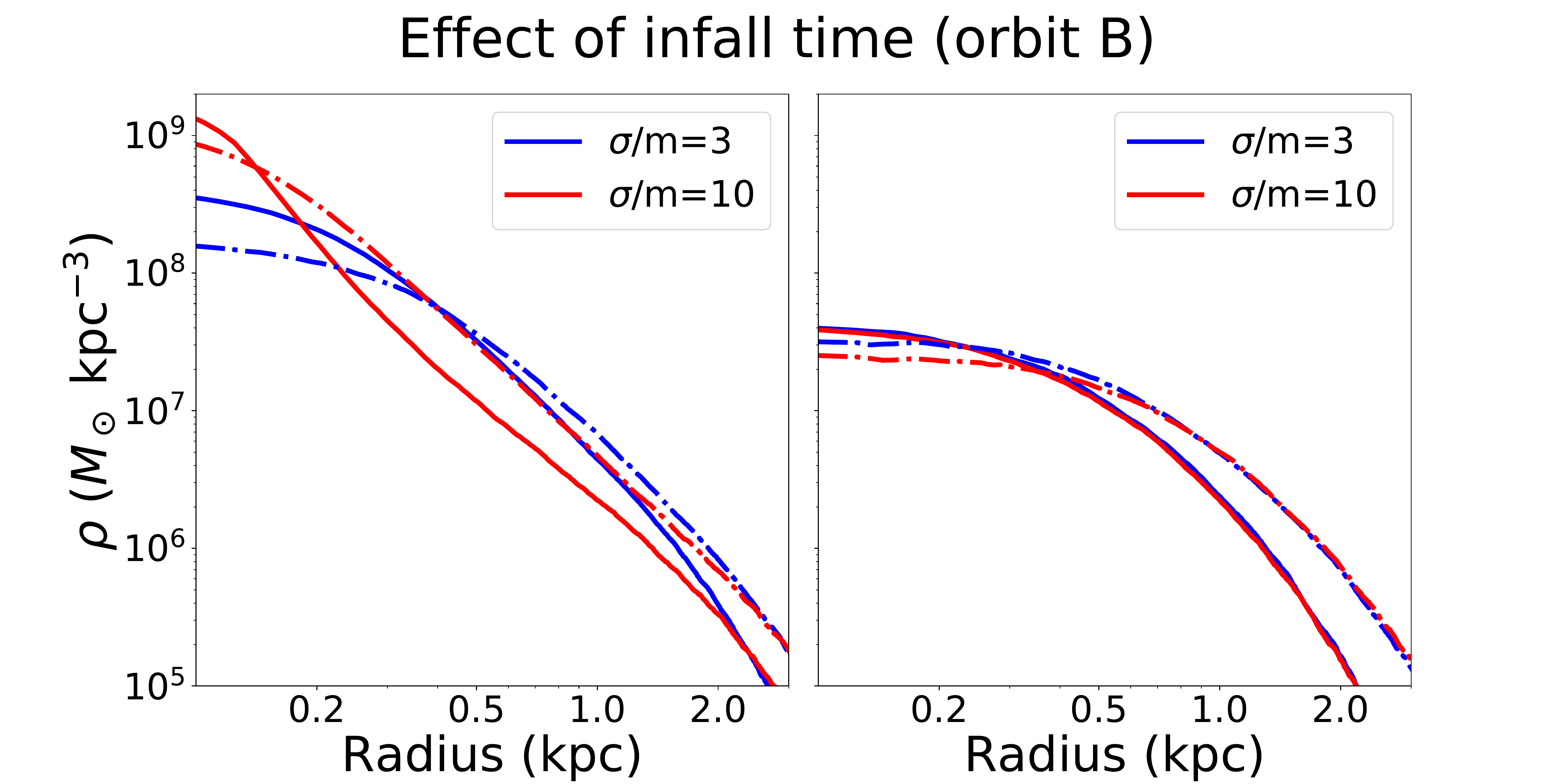}
\caption{The density profile of satellite in a host halo without disk for early infall (solid) and late infall with 5 Gyr of evolution outside the host halo (dot-dashed), in both cases with a total of 10 Gyr evolution time. We test high-concentration (left) and low-concentration (right) initial halos, and self-interaction cross sections of $\sigma/m=3,10$ cm$^2$/g. The early-infall low concentration subhalo in orbit A for $\sigma/m = 10 \, \text{cm}^2/\text{g}$ (solid red line in top right panel) is mostly destroyed at late time.}
\label{fig:test3coreprofile}
\end{figure}

We show in Fig.~\ref{fig:test3coreprofile} the effect of taking field halo density profiles at 5 Gyr (as calculated in the first row of Fig.~\ref{fig:isoall}, but at 5 rather than 10 Gyr) as alternative initial conditions, as a proxy for how different infall times of satellite galaxies could affect the subsequent evolution of their density profiles. In the left panels, displaying high-concentration halos, the 5 Gyr input profile is already in the phase of core-collapse, so the truncation of the outer region does not have a particularly large influence on the central density evolution. In the bottom right panel, the similarity between the initial and evolved profiles indicates that (in the absence of a disk) the influence of tides on the central density is small in Orbit B; the evolution is dominated by self-interactions, and is similar to that for a field halo (as in the top panels of Fig.~\ref{fig:isoall}). In the presence of a disk, tides can become more important even for Orbit B, as indicated in the bottom panels of Fig.~\ref{fig:isoall}.

For low-concentration subhalos on Orbit A, an earlier infall time can lead to significantly more depletion of the subhalo. The size of the effect depends on how long the large core suffers from tides. As shown in the top right panel, for larger self-interaction cross sections, the core develops earlier, making the difference of infall time (i.e. the difference between the solid and dashed lines) more significant. Explicitly, the combination of large core and early infall means that the subhalo spends a long period orbiting within the host with a large core, and is more easily destroyed.

\end{appendix}

\end{document}